\documentstyle[twoside]{article}

\catcode`\@=11
\long\def\@makefntext#1{
\protect\noindent \hbox to 3.2pt {\hskip-.9pt
$^{{\eightrm\@thefnmark}}$\hfil}#1\hfill}		

\def\@makefnmark{\hbox to 0pt{$^{\@thefnmark}$\hss}}	

\def\ps@myheadings{\let\@mkboth\@gobbletwo
\def\@oddhead{\hbox{}
\rightmark\hfil\eightrm\thepage}
\def\@oddfoot{}\def\@evenhead{\eightrm\thepage\hfil
\leftmark\hbox{}}\def\@evenfoot{}
\def\sectionmark##1{}\def\subsectionmark##1{}}

\oddsidemargin=\evensidemargin
\newcounter{sectionc}\newcounter{subsectionc}\newcounter{subsubsectionc}
\renewcommand{\section}[1] {\vspace{12pt}\addtocounter{sectionc}{1}
\setcounter{subsectionc}{0}\setcounter{subsubsectionc}{0}\noindent
	{\tenbf\thesectionc. #1}\par\vspace{5pt}}
\renewcommand{\subsection}[1] {\vspace{12pt}\addtocounter{subsectionc}{1}
	\setcounter{subsubsectionc}{0}\noindent
	{\bf\thesectionc.\thesubsectionc. {\kern1pt \bfit #1}}\par\vspace{5pt}}
\renewcommand{\subsubsection}[1] {\vspace{12pt}\addtocounter{subsubsectionc}{1}
	\noindent{\tenrm\thesectionc.\thesubsectionc.\thesubsubsectionc.
	{\kern1pt \tenit #1}}\par\vspace{5pt}}
\newcommand{\nonumsection}[1] {\vspace{12pt}\noindent{\tenbf #1}
	\par\vspace{5pt}}

\newcounter{appendixc}
\newcounter{subappendixc}[appendixc]
\newcounter{subsubappendixc}[subappendixc]
\renewcommand{\thesubappendixc}{\Alph{appendixc}.\arabic{subappendixc}}
\renewcommand{\thesubsubappendixc}
	{\Alph{appendixc}.\arabic{subappendixc}.\arabic{subsubappendixc}}

\renewcommand{\appendix}[1] {\vspace{12pt}
        \refstepcounter{appendixc}
        \setcounter{figure}{0}
        \setcounter{table}{0}
        \setcounter{lemma}{0}
        \setcounter{theorem}{0}
        \setcounter{corollary}{0}
        \setcounter{definition}{0}
        \setcounter{equation}{0}
        \renewcommand{\thefigure}{\Alph{appendixc}.\arabic{figure}}
        \renewcommand{\thetable}{\Alph{appendixc}.\arabic{table}}
        \renewcommand{\theappendixc}{\Alph{appendixc}}
        \renewcommand{\thelemma}{\Alph{appendixc}.\arabic{lemma}}
        \renewcommand{\thetheorem}{\Alph{appendixc}.\arabic{theorem}}
        \renewcommand{\thedefinition}{\Alph{appendixc}.\arabic{definition}}
        \renewcommand{\thecorollary}{\Alph{appendixc}.\arabic{corollary}}
        \renewcommand{\theequation}{\Alph{appendixc}.\arabic{equation}}
        \noindent{\tenbf Appendix \theappendixc #1}\par\vspace{5pt}}
\newcommand{\subappendix}[1] {\vspace{12pt}
        \refstepcounter{subappendixc}
        \noindent{\bf Appendix \thesubappendixc. {\kern1pt \bfit #1}}
	\par\vspace{5pt}}
\newcommand{\subsubappendix}[1] {\vspace{12pt}
        \refstepcounter{subsubappendixc}
        \noindent{\rm Appendix \thesubsubappendixc. {\kern1pt \tenit #1}}
	\par\vspace{5pt}}

\newcommand{\rmlineskip}{\baselineskip=13pt}
\newcommand{\smalllineskip}{\baselineskip=10pt}

\def\eightcirc{
\begin{picture}(0,0)
\put(4.4,1.8){\circle{6.5}}
\end{picture}}
\def\eightcopyright{\eightcirc\kern2.7pt\hbox{\eightrm c}}

\def\abstracts#1#2#3{{
	\centering{\begin{minipage}{4.5in}\baselineskip=10pt\footnotesize
	\parindent=0pt #1\par
	\parindent=15pt #2\par
	\parindent=15pt #3
	\end{minipage}}\par}}

\renewenvironment{thebibliography}[1]
	{\frenchspacing
	 \ninerm\baselineskip=11pt
	 \begin{list}{\arabic{enumi}.}
	{\usecounter{enumi}\setlength{\parsep}{0pt}
	 \setlength{\leftmargin 12.7pt}{\rightmargin 0pt} 
	 \setlength{\itemsep}{0pt} \settowidth
	{\labelwidth}{#1.}\sloppy}}{\end{list}}

\newcounter{itemlistc}
\newcounter{romanlistc}
\newcounter{alphlistc}
\newcounter{arabiclistc}

\newcommand{\fcaption}[1]{
        \refstepcounter{figure}
        \setbox\@tempboxa = \hbox{\footnotesize Fig.~\thefigure. #1}
        \ifdim \wd\@tempboxa > 5in
           {\begin{center}
        \parbox{5in}{\footnotesize\smalllineskip Fig.~\thefigure. #1}
            \end{center}}
        \else
             {\begin{center}
             {\footnotesize Fig.~\thefigure. #1}
              \end{center}}
        \fi}

\newcommand{\tcaption}[1]{
        \refstepcounter{table}
        \setbox\@tempboxa = \hbox{\footnotesize Table~\thetable. #1}
        \ifdim \wd\@tempboxa > 5in
           {\begin{center}
        \parbox{5in}{\footnotesize\smalllineskip Table~\thetable. #1}
            \end{center}}
        \else
             {\begin{center}
             {\footnotesize Table~\thetable. #1}
              \end{center}}
        \fi}

\def\@citex[#1]#2{\if@filesw\immediate\write\@auxout
	{\string\citation{#2}}\fi
\def\@citea{}\@cite{\@for\@citeb:=#2\do
	{\@citea\def\@citea{,}\@ifundefined
	{b@\@citeb}{{\bf ?}\@warning
	{Citation `\@citeb' on page \thepage \space undefined}}
	{\csname b@\@citeb\endcsname}}}{#1}}

\newif\if@cghi
\def\cite{\@cghitrue\@ifnextchar [{\@tempswatrue
	\@citex}{\@tempswafalse\@citex[]}}
\def\citelow{\@cghifalse\@ifnextchar [{\@tempswatrue
	\@citex}{\@tempswafalse\@citex[]}}
\def\@cite#1#2{{$\null^{#1}$\if@tempswa\typeout
	{IJCGA warning: optional citation argument
	ignored: `#2'} \fi}}

\def\pmb#1{\setbox0=\hbox{#1}
	\kern-.025em\copy0\kern-\wd0
	\kern.05em\copy0\kern-\wd0
	\kern-.025em\raise.0433em\box0}

\def\fnt#1#2{\footnotetext{\kern-.3em
	{$^{\mbox{\scriptsize #1}}$}{#2}}}

\def\fpage#1{\begingroup
\voffset=.3in
\thispagestyle{empty}\begin{table}[b]\centerline{\footnotesize #1}
	\end{table}\endgroup}

\def\runninghead#1#2{\pagestyle{myheadings}
\markboth{{\protect\footnotesize\it{\quad #1}}\hfill}
{\hfill{\protect\footnotesize\it{#2\quad}}}}
\font\tenrm=cmr10
\font\tenit=cmti10
\font\tenbf=cmbx10
\font\bfit=cmbxti10 at 10pt
\font\ninerm=cmr9

\font\eightrm=cmr8

\def\qed{\hbox{${\vcenter{\vbox{			
   \hrule height 0.4pt\hbox{\vrule width 0.4pt height 6pt
   \kern5pt\vrule width 0.4pt}\hrule height 0.4pt}}}$}}

\newcommand{\Det}{{\rm Det}}
\newcommand{\Tr}{{\rm Tr}}
\newcommand{\Log}{{\rm Log}}
\newcommand{\Dslash}{{D \hskip -6pt /}}

\begin{document}

\runninghead{G. Dunne} {An All-Orders Derivative Expansion}

\normalsize\rmlineskip
\thispagestyle{empty}
\setcounter{page}{1}


\vspace*{0.88truein}

\fpage{1}
\centerline{\bf AN ALL-ORDERS DERIVATIVE EXPANSION\footnote{Talk given at the
Telluride Workshop on Low Dimensional Field Theory, August 1996.}}
\vspace*{0.37truein}
\centerline{\footnotesize Gerald DUNNE}
\vspace*{0.015truein}
\centerline{\footnotesize\it Physics Department}
\baselineskip=10pt
\centerline{\footnotesize\it University of Connecticut, Storrs, CT 06269}
\vspace*{10pt}
\vspace*{0.225truein}

\vspace*{0.21truein}
\abstracts{We evaluate the exact ${\rm QED}_{2+1}$ effective action for
fermions in the presence of a family of static but spatially inhomogeneous
magnetic field profiles. This exact result yields an all-orders derivative
expansion of the effective action, and indicates that the derivative expansion
is an asymptotic, rather than a convergent, expansion. }{}{}

\rmlineskip			
\vspace*{12pt}			

\vspace*{-0.5pt}
\noindent
The effective action is a fundamental tool for the study of quantum field
theory. Using the proper-time technique, Schwinger\cite{schwinger} showed that
the QED effective action can be computed exactly for either a {\it constant}
or a {\it plane wave} electromagnetic field. This result was later
adapted to ${\rm QED}_{2+1}$ with constant fields by Redlich\cite{redlich}. For
general electromagnetic fields the effective action cannot be computed exactly,
so one must resort to some sort of perturbative expansion. A common approach is
the derivative expansion\cite{aitchison,lee} in which the field is assumed to
vary only very `slowly'. A useful computational procedure for the derivative
expansion\cite{cangemi,shovkovy} is provided by the supersymmetric path
integral representation of the proper-time effective action\cite{rajeev}.
However, even first-order derivative expansion calculations of the effective
action are rather cumbersome, and it is very unclear just how convergent (or
otherwise) the expansion is. Thus, the physical significance of a first-order
correction term must be examined with care. In this paper we seek a better
nonperturbative understanding of the derivative expansion by considering a new
exactly solvable model which has a spatially varying magnetic field. The {\it
exact} integral representation for the effective action may then be expanded in
an {\it asymptotic series}, and we find that the first few terms agree with
(independent) derivative expansion computations.

Specifically, we show that the ${\rm QED}_{2+1}$ effective action for massive
fermions can be computed exactly\cite{cangemi2} in the presence of
time-independent but
spatially varying magnetic fields of the form:
\begin{equation}
  B(x,y) = B\,{\rm sech}^2 (\frac{x}{\lambda})
\label{mag}
\end{equation}
Here, the magnetic field is uniform in the $y$-direction but is inhomogeneous
in the $x$-direction. The length scale $\lambda$ characterizes the width of the
profile in the $x$-direction. In the limit $\lambda \to \infty$ the magnetic
field in (\ref{mag}) tends to a uniform field of strength $B$. The constant $B$
(the peak value of the magnetic field) sets a length scale $1/\sqrt{B}$ known
as the magnetic length, and the derivative expansion regime corresponds to
$\lambda\gg 1/\sqrt{B}$. In practice, the system will be considered in a box in
the $y$-direction of size $L$. The total magnetic flux $\Phi$ is then finite:
$\Phi = B\lambda L/\pi$. We show that the effective action has a simple {\it
exact} integral representation involving elementary functions for all values of
mass $m$, width $\lambda$, and strength $B$.

The fermionic effective action is given by
\begin{equation}
  i S_{\rm eff}=\Log\, \Det \left(i\Dslash -m \right)
\label{eff}
\end{equation}
where $\Dslash = \gamma^\nu D_\nu =\gamma^\nu
\left(\partial_\nu+iA_\nu\right)$, and $A_\nu$ is a fixed classical gauge
potential. The Minkowski gamma matrices $\gamma^\nu$ satisfy the
anticommutation relations $\{ \gamma^\nu , \gamma^\sigma \}= -2 g^{\nu\sigma}=2
\, {\rm diag}(1,-1,-1)$, and in $2+1$ dimensions the irreducible gamma matrices
may be chosen as the $2\times 2$ matrices:
$\gamma^0=\sigma^3,\, \gamma^1=i\sigma^1,\,\gamma^2=i\sigma^2$, where the
$\sigma^i$ are the $2\times 2$ Pauli matrices. Note that an
alternative choice, $\gamma^0=-\sigma^3$, $\gamma^1=-i\sigma^1$,
$\gamma^2=-i\sigma^2$, can be viewed as changing the sign of the mass,
$m\to-m$, in the Dirac operator. The system with effective action~(\ref{eff})
is not parity invariant since a fermion mass term breaks parity in $2+1$
dimensions \cite{parity}. However, in this paper we shall limit our attention
to the parity-even part of the effective action (\ref{eff}):
\begin{eqnarray}
  i S=\frac{1}{2}\Log\,\Det \left[\left(i\Dslash - m\right)\left(-i\Dslash -
m\right)\right]=\frac{1}{2}\Tr\,\Log \left(\Dslash^2+m^2 \right)
\label{eff-2}
\end{eqnarray}
The $\Tr\,\Log$ may be defined by a ``proper-time'' integral
representation\cite{schwinger}
\begin{equation}
i S=-\frac{1}{2}\int_0^\infty\,{ds\over s}\,e^{-m^2
s}\,\Tr\left(e^{-s(D\hskip-5pt /)^2}\right)
\label{proper}
\end{equation}
{}From this expression we see that in order to compute the effective action in
this proper-time form we need to compute the `proper-time propagator'
\begin{equation}
\Tr\left(e^{-s(D\hskip-5pt /)^2}\right)
\label{propagator}
\end{equation}
When the field strength $F_{\mu\nu}$ is constant, this propagator may be
computed exactly because the Fock-Schwinger gauge choice,
$A_\mu=\frac{1}{2}x^\nu F_{\nu\mu}$, leads to $(\Dslash)^2$ being a quadratic
operator. Thus the propagator in (\ref{propagator}) is essentially that of a
harmonic oscillator, leading to the simple expression\cite{redlich}
\begin{equation}
S=-{{\cal A}\over 8\pi^{3/2}}\int_0^\infty\,{ds\over s^{3/2}}\,e^{-m^2
s}\,\left[ |{\cal F}|\coth\left(|{\cal F}| s\right)-{1\over s}\right]
\label{red}
\end{equation}
where $|{\cal F}|=\sqrt{F_{\mu\nu}F^{\mu\nu}}$ and ${\cal A}$ is an area factor
associated with the degeneracy. Note that the $1/s$ term in the square
parentheses in (\ref{red}) corresponds to the zero field case; thus, one
actually computes the {\it difference} between the effective action with and
without the background field.

Ideally we would like to go beyond this constant field case and compute the
effective action in an {\it arbitrary} background electromagnetic field.
However, as this is not feasible, we consider a derivative
expansion\cite{aitchison,lee}, which is a perturbative expansion about the
constant field case. Consider, for example, a Taylor expansion of the gauge
field
\begin{equation}
A_\mu =\frac{1}{2}x^\alpha F_{\alpha\mu}+\frac{1}{3}x^\alpha x^\beta
\partial_\alpha F_{\beta\mu}+\dots
\label{derivative}
\end{equation}
Here the derivative terms $\partial F$ are in some sense smaller (see below)
than the $F$ terms. The operator $\Dslash^2$ now involves terms {\it quartic}
in $x$, so the proper-time propagator (\ref{propagator}) can be computed, at
best, in perturbation theory. An efficient way to compute these perturbative
contributions is to re-express the proper-time integral representation of the
effective action as a SUSY quantum mechanical path
integral\cite{cangemi,shovkovy}. However, there is a proliferation of terms
appearing in the perturbation expansion. For example, in the $2+1$ dimensional
case there are $6$ different terms\footnote{Various inequivalent contractions
of the tensors $F_{\mu\nu}$, $\partial_\alpha F_{\beta\mu}$, $g_{\mu\nu}$ and
$\epsilon_{\alpha\beta\gamma}$ appear. In $3+1$-dimensions there are many more
terms\cite{lee} due to the epsilon tensor
$\epsilon_{\alpha\beta\gamma\delta}$.} ~in the first order correction to the
constant field case. For definiteness we choose one such term by restricting to
a static, but spatially inhomogeneous, magnetic field (no electric field). Then
the first order correction to the effective action is \cite{cangemi}
\begin{equation}
\Delta S= -\frac{1}{8}\left({1\over 4\pi}\right)^{3/2} \int d^3x
{\left(\vec{\nabla} B\right)^2 \over B^{3/2}}\int_0^\infty {ds\over s^{1/2}}\,
e^{-m^2s/B} \left({d\over ds}\right)^3\left[s\,coth(s)\right]
\end{equation}
For $m\ll B/m$ (i.e. for the fermion rest mass energy scale much smaller than
the cyclotron energy scale set by the approximately uniform magnetic
background) this correction simplifies to
\begin{equation}
\Delta S={1\over \sqrt{2}(4\pi)^2} {15\over 16\pi} \zeta\left(5/2\right) \int
d^3x{(\vec{\nabla}B)^2\over B^{3/2}}
\label{first}
\end{equation}
where $\zeta(z)=\sum_{n=1}^\infty n^{-z}$ is the Riemann zeta function and
$\zeta(5/2)\approx 1.34$.

Computing higher order terms in the derivative expansion is messy and not
particularly illuminating. Thus, to obtain a better understanding of the
meaning of the expansion we seek an {\it exactly solvable} effective action
with an inhomogeneous background field. To be specific, we consider static
magnetic fields. Then the proper-time expression (\ref{proper}) becomes
\begin{equation}
S=-\frac{1}{4\sqrt{\pi}}\sum_{\pm}\int_0^\infty \, {ds\over s^{3/2}}
\Tr\left(e^{-(m^2-\vec{D}^2\pm B)s}\right)
\label{pt}
\end{equation}
after taking the trace over the energy. We must also sum over the two spin
projections $\pm$ arising in the Pauli operator $m^2-\vec{D}^2\pm B$, where
$B=F_{12}$ is the magnetic field strength.

An alternative expression for the static effective action is obtained by
performing the proper time integral in (\ref{pt}), yielding
\begin{equation}
S=\frac{1}{2}\sum_{\pm}\Tr\sqrt{m^2-\vec{D}^2\pm B}
\label{vacuum}
\end{equation}
which expresses the static effective action as a sum over the Dirac sea
energies for single fermion states\cite{salam} (recall that the static
effective action is {\it minus} the static effective energy).

Yet another expression for the static effective action is the ``resolvent
form''
\begin{equation}
S=\sum_{\pm}\int {dE\over 2\pi i}\, E^2 \,\Tr\left({1\over m^2-\vec{D}^2\pm
B-E^2}\right)
\label{resolvent}
\end{equation}
which reduces to the Dirac sea form (\ref{vacuum}) after integration over an
appropriate energy contour, and which may also be derived formally from the
original expression $iS=\frac{1}{2}\Tr\Log(\Dslash^2+m^2)$ after an integration
by parts in the energy trace.

The three expressions (\ref{pt}), (\ref{vacuum}), and (\ref{resolvent}) are
equivalent but one may be more useful than the others, depending on the form of
the Pauli operator. For example, consider the uniform magnetic field case. The
static vector potential can be chosen as $\vec{A}=(0,B x)$, in which case the
Pauli operator becomes
\begin{equation}
m^2-\vec{D}^2\pm B=m^2-\left({d\over dx}\right)^2+B^2\left(x-{k\over
B}\right)^2\pm B
\label{ho-pauli}
\end{equation}
where $k$ is a momentum variable corresponding to the free $y$ direction. This
Pauli operator has the form of a 1-dimensional harmonic oscillator with
frequency $\omega=2B$, and thus has a purely discrete spectrum of levels known
as Landau levels, each of which is infinitely degenerate. In fact, the
degeneracy is given by the (infinite) total flux, which in a finite area ${\cal
A}$ is $B{\cal A}/2\pi$. Thus the proper time propagator (\ref{propagator}) is
\begin{equation}
\sum_\pm\Tr\left(e^{-(m^2-\vec{D}^2\pm B)s}\right) = {B{\cal A}\over
2\pi}e^{-m^2s}\sum_\pm\sum_{n=0}^\infty e^{-2B(n+\frac{1}{2}\pm \frac{1}{2})s}=
{B{\cal A}\over 2\pi} \coth(Bs)\, e^{-m^2 s}
\label{pt-ho}
\end{equation}
which agrees with Redlich's result (\ref{red}). The sum over Dirac energies in
(\ref{vacuum}) may also be performed explicitly because the spectrum is
discrete:
\begin{eqnarray}
\frac{1}{2}\sum_\pm\Tr\sqrt{m^2-\vec{D}^2\pm B}&=&{B{\cal A}\over 4\pi}\sum_\pm
\sum_{n=0}\sqrt{m^2+2B(n+1/2)\pm B}\nonumber\\
&=& {B^{3/2}{\cal A}\over 2\sqrt{2}\pi}\left[\frac{m}{\sqrt{2B}}+2
\zeta\left(-\frac{1}{2},1+\frac{m^2}{2B}\right)\right]
\label{hurwitz}
\end{eqnarray}
where $\zeta(z,a)=\sum_{n=0}^\infty (n+a)^{-z}$ is the Hurwitz Zeta
function\cite{bateman}. Using the integral representation\cite{grad}
\begin{equation}
\Gamma(z)\left[\zeta(z,a+1)+\frac{a^{-z}}{2}\right]=\int_0^\infty ds
\,e^{-2as}\,(2s)^{z-1}  \coth\,s
\end{equation}
we see that the proper time expression (\ref{red}) provides an integral
representation of the Zeta function expression (\ref{hurwitz}).

These two approaches [the proper-time method (\ref{pt}) and the Dirac sea
method (\ref{vacuum})] are the most direct, and most familiar, ways to compute
the effective action with a constant $B$ field. However, in the case of an
inhomogeneous magnetic field considered below the resolvent method proves more
useful. Thus we first review the resolvent approach to the constant field case.

With the vector potential $\vec{A}=(0,B x)$, the static effective action
(\ref{resolvent}) is
\begin{equation}
S=\sum_\pm \int{dk\over 2\pi}\int {dE \over 2\pi i} \, E^2 \,\Tr\,
G_{E^2,k}^{(\pm)}
\label{ho-eff}
\end{equation}
Here $G_{E^2,k}^{(\pm)}$ is the resolvent Green's function for the
one-dimensional Schr\"odinger operator in (\ref{ho-pauli}).
The diagonal resolvent may be constructed, as usual, as the product of two
independent solutions $u_1(x)$ and $u_2(x)$ to the corresponding homogeneous
differential equation, divided by their Wronskian:
\begin{equation}
G(x,x)={u_1(x) u_2(x)\over W}
\end{equation}
The Wronskian $W=u_1^\prime(x) u_2(x)-u_1(x) u_2^\prime (x)$ is independent of
$x$. For the harmonic oscillator operator in (\ref{ho-pauli}) the independent
solutions are parabolic cylinder functions $D_\nu(x)$, and so the trace of the
diagonal resolvent gives (with a straightforward regularization)
\begin{equation}
\int{dk\over 2\pi}\,\Tr\, G_{E^2,k}^{(\pm)}= {{\cal A}\over
4\pi}\,\psi(-\nu_\pm)
\label{resolvent-ho}
\end{equation}
where $\nu_\pm=(E^2-m^2-B(1\pm 1))/(2B)$, and $\psi(z)=(d/dz)\log\Gamma(z)$ is
the Euler Psi function\cite{bateman}. The Psi function has simple poles at the
negative integers (corresponding to the familiar discrete spectrum of the HO
system), and so the $E$ integral in (\ref{ho-eff}) may be performed to give the
Zeta function expression in (\ref{hurwitz}).

To go beyond the constant field case, we choose a vector potential of the form
$\vec{A}=(0,a(x))$, which corresponds to a magnetic field $B(x)=a^\prime(x)$
only depending on the $x$-coordinate. The Pauli operator for such a vector
potential is
\begin{equation}
m^2-\vec{D}^2\pm B(x)=m^2-\left(\frac{d}{dx}\pm [a(x)-k]\right)
\left(\frac{d}{dx}\mp [a(x)-k]\right)
\label{susy}
\end{equation}
which has the form of a one-dimensional\footnote{Note, however, that the {\it
superpotential}, $Q_k(x)=a(x)-k$, is $k$ dependent, where $k$ is the momentum
for the free $y$-direction. This reflects the fact that the Pauli operator is
really a two-dimensional operator.}~ SUSY quantum mechanics
operator\cite{witten}. This special form for the vector potential is chosen
because for various particular choices of the {\it superpotential},
$Q_k(x)=a(x)-k$, the SUSY Pauli operator has an exactly known spectrum. Given
complete knowledge of the spectrum it should then be possible, at least in
principle, to compute the effective action using one of the three expressions
(\ref{pt}), (\ref{vacuum}), or (\ref{resolvent}). With $Q_k(x)=Bx-k$ we obtain
the constant field case discussed above. With the superpotential choice
\begin{equation}
Q_k(x)=B\lambda \tanh(\frac{x}{\lambda})-k
\end{equation}
the corresponding magnetic field profile is as in (\ref{mag}), and the Pauli
operator has an exactly solvable spectrum\cite{morse}.

However, there are two crucial differences from the constant field case. First,
the spectrum depends {\it explicitly} on the momentum parameter $k$. Therefore,
the trace over $k$ does not simply lead to an overall flux factor [as was the
case in (\ref{pt-ho}), (\ref{hurwitz}) and \ref{resolvent-ho})], but must be
truly integrated over. Second, the spectrum is no longer purely discrete - it
has continuum states (both one-way and two-way scattering states) as well as
discrete bound states. Moreover, the density of states depends explicitly on
$k$, so that, for example, the actual {\it number} of bound states depends on
$k$. Thus, an explicit sum over the spectrum as in (\ref{vacuum}) is
problematical, and it is nontrivial to find a simple representation of the
proper time propagator in (\ref{pt}). Nevertheless, it is possible to find the
trace of the diagonal resolvent in (\ref{resolvent}).

The corresponding homogeneous differential equation can be converted to
hypergeometric form with the substitution $\xi=[1+tanh(x/\lambda)]/2$. The two
independent solutions may be chosen as (here $\pm$ refers to the two spin
projections in (\ref{susy}))
\begin{eqnarray}
& u_1(x) = \xi^\alpha (1-\xi)^\beta F(\alpha+\beta\mp B\lambda^2
,\alpha+\beta+1\pm B\lambda^2;1+2\alpha;\xi) \\[6pt]
& u_2(x) = \xi ^\alpha (1-\xi)^\beta F(\alpha+\beta\mp B\lambda^2
,\alpha+\beta+1\pm B\lambda^2;1+2\beta;1-\xi)
\end{eqnarray}
Here, $\alpha $ and $\beta$ are defined as the following roots (with positive
real part):
\begin{equation}
  \alpha = \frac{\lambda }{2}\sqrt{\alpha_{-k}^2 - E^2} \hspace{34pt}
  \beta = \frac{\lambda }{2}\sqrt{\alpha _{k}^2 - E^2} \hspace{34pt}
  \alpha_k = \sqrt{(k - B \lambda)^2 + m^2}
\end{equation}
With these conventions, $u_1$ is regular at $\xi=0$ (i.e. as $x \to
-\infty$), whereas $u_2$ is regular at $\xi=1$ (i.e. as $x \to
+\infty$).  The Wronskian of these solutions is a constant:
\begin{equation}
  W = u_1'(x) u_2(x) - u_1(x) u'_2(x) = \frac{2}{\lambda}
  \frac{\Gamma(1+2\alpha) \Gamma(1+2\beta)}{\Gamma(\alpha+\beta\mp B\lambda^2)
\Gamma(\alpha+\beta+1\pm B\lambda^2)}
\end{equation}
The trace of the diagonal resolvent is then
\begin{equation}
\hspace{-4pt}\int dx ~G_{E^2,k}^{(\pm)} (x,x) =
- \frac{\lambda^2}{4} \left(\frac{1}{\alpha} +
    \frac{1}{\beta} \right) \left[ \psi(\alpha+\beta\mp B\lambda^2) +
    \psi(\alpha+\beta+1\pm B\lambda^2) \right]
\label{trace}
\end{equation}
The spectrum contains a finite number of bound states, which arise from the
(simple) poles of the $\psi$-functions in (\ref{trace}). The same discrete
spectrum may be obtained by solving the homogeneous equation for real $E^2$ and
requiring normalizability of the eigenfunctions\cite{morse}. The spectrum also
contains a cut starting at $\alpha_k^2$ and another cut starting at
$\alpha _{-k}^2$, corresponding to the two barrier thresholds of the
`potential' $m^2+Q_k^2\pm Q_k^\prime$ in (\ref{susy}).

Remarkably, the remaining integral over $k$ may be performed\cite{cangemi2},
leaving an integral form for the effective action:
\begin{equation}
  S = \frac{L}{4\pi\lambda^2} \sum_\pm\int_{z_0}^\infty dz\;
  \frac{z (z^2 - z_0^2)}{\sqrt{z^2 - z_0^2 + (\lambda m)^2}} \; \bigl
  [ \psi(z\mp B\lambda^2) + \psi(z+1\pm B\lambda^2)
\bigr]
\label{simple}
\end{equation}
where $ z_0 \equiv \lambda \sqrt{(B\lambda )^2 + m^2}$. This
expression can be rewritten in terms of elementary integrals using the
following representation of the $\psi$-function\cite{bateman}:
\begin{equation}
  \psi (x) = \ln x -\frac{ 1}{2x} -2 \int _0 ^\infty
  \frac{t~dt}{(t^2+x^2)(e^{2\pi t} -1)}
\label{psifunction}
\end{equation}
After summing over both spin projections in (\ref{simple}), we find that the
first two terms on the right hand side in (\ref{psifunction}) contribute
$B$-independent terms to the effective action and therefore may be dropped. For
the third term, the $z$-integral in (\ref{simple}) can be carried out exactly
and we obtain the following finite integral representation for the effective
action
\begin{equation}
  S= -\frac{L}{4\pi\lambda^2} \int_
  0 ^\infty \frac{dt}{e^{2 \pi t}- 1} ~\left ( (B\lambda^2-it)
    \frac{(\lambda ^2 m^2+v^2)}{v} \ln \frac{\lambda m
      -iv}{\lambda m +iv} + c.c. \right )
\label{answer}
\end{equation}
where $c.c.$ denotes the complex conjugate, and $v^2=t^2 + 2i\,t\,B\lambda^2$.
Expression~(\ref{answer}) gives an exact integral representation for the
effective action with the inhomogeneous magnetic
background (\ref{mag}), just as (\ref{red}) gives an exact integral
representation for the effective action with a uniform magnetic
background.

In the limit of vanishing mass, $m=0$, the only relevant dimensionless
parameter is $B\lambda^2$, which is the square of the ratio of the width
parameter $\lambda$ to the magnetic length $1/\sqrt{B}$. Since the {\it
uniform} field case corresponds to $\lambda\gg 1/\sqrt{B}$, we expect the
derivative expansion regime to correspond to the regime in which
$B\lambda^2\to\infty$. It is straightforward to develop an asymptotic expansion
of the integral (\ref{answer}) for large $B\lambda^2$ when $m=0$, yielding
\begin{eqnarray}
  S &=& -\frac{L \lambda(B)^{3/2}}{8 \pi}\sum_{j=0}^\infty
  \frac{1}{(4\pi B\lambda^2)^j} \frac{\Gamma(j-3/2)\Gamma(j+5/2)}{
    \Gamma(j+1) \Gamma(j/2+1/4) \Gamma(3/4-j/2)} \zeta(j+3/2)\nonumber \\
&=& -\frac{L \lambda(B)^{3/2}}{8\sqrt{2} \pi}\left[ \zeta(3/2)-
    \frac{15}{ 16\pi}\zeta(5/2) \frac{1}{B\lambda^2}+\dots\right]
\label{all}
\end{eqnarray}
where $\zeta(z)$ is the Riemann Zeta function\cite{bateman}.

We now compare this expansion with the first two terms of the derivative
expansion (with $m=0$) obtained using the proper-time method\cite{cangemi}:
\begin{equation}
S=\int d^3x\left[\frac{\zeta(-1/2)}{\sqrt{2}\pi}[B(x)]^{3/2}+{1\over \sqrt{2}
(4\pi)^2}{15\over 16\pi}\zeta(5/2){[\vec{\nabla}B(x)]^2\over
[B(x)]^{3/2}}+\dots \right]
\end{equation}
The first term comes from (\ref{hurwitz}) and the second from (\ref{first}).
For the magnetic field profile $B(x,y)=B\, sech^2(x/\lambda)$ in (\ref{mag}),
the $x$ and $y$ integrals can be done (as everything is static, the overall
time scale has been absorbed into the definition of the energy trace
throughout). Then, using the facts that $\int dx sech ^3x =\pi/2$, $\int dx
sech x \, tanh^2 x=\pi/2$, and $\zeta(3/2)=-4\pi \zeta(-1/2)$, we see that
these first two terms of the derivative expansion agree exactly with the first
two terms of the expansion in (\ref{all}). Thus, it is indeed natural to
identify the expansion in (\ref{all}) as an {\it all-orders} derivative
expansion. This expansion, however, is an {\it asymptotic} expansion rather
than a {\it convergent} one. Indeed, as is typical of asymptotic series, the
magnitude of the expansion coefficients decreases at first, then bottoms out
and eventually tends to infinity. Also note that the sign of the expansion
coefficients follows the pattern $+$, $-$, $-$, $-$, $+$, $+$, $-$, $-$, $+$,
$+$, $\dots$ . The eventual double-plus/double-minus pattern is reminiscent of
Stirling's asymptotic series for the Gamma function.

For nonzero mass $m$, there is another dimensionless parameter,
$B/m^2$, which is the ratio of the cyclotron energy to the rest mass
energy. A double expansion of (\ref{answer}) yields
\begin{equation}
  S = - \frac{L m^3 \lambda}{8 \pi} \sum _{j=0}^\infty
  \frac{(2B\lambda^2)^{-j}}{j!} \sum_{k=1}^\infty
  \frac{(2k+j-1)!}{(2 k)!} \frac{{\cal B}_{2k+2j}}
{(2k+j-\frac{1}{2})(2k+j-\frac{3}{2})}
  \left(\frac{2B}{m^2} \right)^{2k+j}
\label{doublesum}
\end{equation}
with ${\cal B}_{n}$ the $n^{th}$ Bernoulli number\cite{bateman,grad}. Each
power in $1/(B\lambda^2)$ corresponds to a fixed order in the derivative
expansion of the effective action.  The zeroth order term agrees with
the $B/m^2$ expansion of the exact constant $B$ field answer\cite{redlich},
while the $1/(B\lambda^2)$ term agrees with the
$B/m^2$ expansion of the leading derivative expansion
contribution\cite{cangemi}.

I conclude by commenting briefly on some natural extensions of these results.
First, these results could be extended to $3+1$ dimensions; although there one
must be careful to include the proper charge renormalization, which is present
even in the constant field case\cite{schwinger}. Second, it would be
interesting to compute the {\it finite temperature} effective action for this
$QED_{2+1}$ system. If one computes the zero temperature effective action using
the proper time expression (\ref{pt}) then the extension to nonzero temperature
is immediate\cite{cangemi3}:
\begin{equation}
  S_{\rm eff}=- {1\over 4\sqrt{\pi}}\sum_\pm \int_0^\infty{ds\over s^{3/2}} \;
  \theta_4{\left({i\beta \mu \over 2}\left| { i \beta^2 \over 4\pi
          s}\right.\right)} \Tr\left[ \exp-\left(m^2-\vec{D}^2\pm
B\right)s\right]
\label{th4}
\end{equation}
Here $\beta$ is the inverse temperature, $\mu$ is the chemical potential and
the fourth theta function is\cite{grad}
\begin{equation}
  \theta_4 {\left(u | \tau\right)} =1+2\sum_{n=1}^\infty (-1)^n
  e^{i\pi\tau n^2} \cos\left(2nu\right)
\label{theta4}
\end{equation}
In the zero chemical potential and zero temperature limit, one simply omits the
theta factor in the proper-time integral, giving (\ref{pt}). However, since
this theta factor is independent of the magnetic field, if one had computed the
proper-time propagator for this system (rather than using the resolvent
technique described in this paper) then the exact finite temperature effective
action would follow immediately. It would be interesting to investigate the
properties of the derivative expansion at finite temperature using this exactly
solvable model.

\vskip 1in

\nonumsection{Acknowledgments}
\noindent  This talk is based on work done in collaboration with Daniel Cangemi
and Eric D'Hoker, whom I thank for discussions and correspondence. This
research was supported by the Department of Energy under Grant
No.~DE-FG02-92ER40716.00 and by the University of Connecticut Research
Foundation.

\end{document}